\RequirePackage[2020-02-02]{latexrelease}
\documentclass[floats,aps,onecolumn,secnumarabic,tightenlines,nofootinbib]{revtex4}
\usepackage{amssymb,graphics,graphicx,amstext,amsmath,amsfonts,physics,bm,color,appendix}

\begin{document}


\title{Non-local gravity effects in cosmological dynamics\\probed by IceCube/KM3NeT signals and dark matter relic abundance}

\author{S.\ Capozziello$^{abc}$, G.\ Lambiase$^{de}$ and G.\ Meluccio$^{bcf}$}

\affiliation{$^a$Dipartimento di Fisica  ``E.\ Pancini'', Università degli Studi di Napoli ``Federico II'', Compl.\ Univ.\ di Monte S.\ Angelo, Edificio G, Via Cinthia, I-80126, Napoli, Italy}
\affiliation{$^b$Istituto Nazionale di Fisica Nucleare (INFN), Sez.\ di Napoli, Compl.\ Univ.\ di Monte S.\ Angelo, Edificio G, Via	Cinthia, I-80126, Napoli, Italy}
\affiliation{$^c$Scuola Superiore Meridionale, Via Mezzocannone 4, I-80134, Napoli, Italy}
\affiliation{$^d$Dipartimento di Fisica ``E.\ R.\ Caianiello'' Università di Salerno, I-84084 Fisciano (SA), Italy}
\affiliation{$^e$INFN -- Sez.\ di Napoli, Gruppo Collegato di Salerno, Italy}
\affiliation{$^f$Department of Physics, Loyola Marymount University, 1 LMU Drive, 90045, Los Angeles, California, USA}

\date{\today}

\def\be{\begin{equation}}
\def\ee{\end{equation}}
\def\al{\alpha}
\def\bea{\begin{eqnarray}}
\def\eea{\end{eqnarray}}
\renewcommand{\theequation}{\thesection.\arabic{equation}}

\begin{abstract}
Non-local gravity terms have a relevant role in determining the  cosmological dynamics. Here we consider curvature- and torsion-based cosmological models where non-local terms can be ``scalarised'' and then reduced under the standard of scalar-tensor gravity. In this context, we study the role of non-local cosmology with regards to the recent results reported by the IceCube/KM3NeT experiments, which revealed high-energy astrophysical neutrino fluxes up to energies of $220$\,PeV. Specifically, we consider the four-dimensional operator $y_{\alpha\chi}\bar L_\alpha H\chi$ in order to explain both the neutrino rate result and the abundance of dark matter in the Universe, provided that the cosmological background evolves according to non-local gravitational field equations. We show that different dynamical systems representing the evolution of the Universe can be highly sensitive to the parameters of non-local gravity at energies probed by IceCube/KM3NeT. In particular, we adopt power law solutions inferred by the existence of Noether symmetries in non-local cosmological models.
\\
\\

\end{abstract}

\maketitle


\section{Introduction}

General Relativity (GR) currently provides the most accurate description of gravitational interactions. A vast body of observations, ranging from the Solar System to cosmological scales, has repeatedly confirmed its validity. This includes, for instance, the observed expansion of the Universe and the formation of large-scale structures, both extensively probed at cosmological as well as astrophysical scales. Furthermore, the recent detection of gravitational waves and the observational evidence of black holes represent some of the most compelling confirmations of Einstein's theory.

Despite these remarkable successes, GR exhibits several shortcomings, arising at ultraviolet (UV) and infrared (IR) scales. These issues suggest that GR should be regarded as an effective field theory, valid  within certain energy regimes, but not capable of representing the ultimate theory of gravity at all scales. The shortcomings are of different nature. From a mathematical standpoint, they manifest as well-known spacetime singularities. For example, the Kruskal--Szekeres maximal extension of the Schwarzschild solution reveals a genuine spacetime singularity at the gravitational centre, which cannot be removed by any coordinate transformation.

At astrophysical scales, one of the most prominent problems is the discrepancy between theoretical predictions and the observed velocities of stars orbiting the centres of disk galaxies, known as the ``galaxy rotation curve problem'' \cite{Babocock,Bosma:1981zz}. To address this issue, a hypothetical perfect fluid -- referred to as dark matter (DM) -- is introduced, which is expected to account for approximately 26$\%$ of the total energy content of the Universe.

At cosmological scales, current observations point out that the present-day Universe is undergoing an accelerated phase of expansion. This phenomenon can be explained by introducing the cosmological constant $\Lambda$, which corresponds to a perfect fluid with negative pressure. Its dynamical counterpart is commonly referred to as dark energy, capable of driving the accelerated cosmic expansion \cite{Peebles:2002gy,Padmanabhan:2002ji}. However, there is currently no direct experimental evidence for the existence of either dark matter or dark energy, nor a clear understanding of their physical nature.\footnote{Although the $\Lambda$CDM model of cosmology is consistent with observations \cite{Spergel:2003cb}, it suffers from several conceptual issues, such as the discrepancy between the observed value of the energy density associated with $\Lambda$ and the vacuum energy density predicted by Quantum Field Theory (QFT), which differ by approximately 120 orders of magnitude.}

Significant obstacles arise in the attempt to reconcile GR with QFT. In the UV regime, GR is perturbatively non-renormalizable, as divergences persist beyond the one-loop level \cite{Goroff:1985th}. Moreover, GR cannot be formulated within the Yang--Mills framework, which has thus far precluded its consistent unification with the other fundamental interactions. Further conceptual issues include the absence of a well-defined Hilbert space for gravitational systems and the lack of a clear probabilistic interpretation of the corresponding wave function. As a result, a complete and consistent quantum theory of gravity remains elusive. Extended theories of gravity (ETGs) provide a possible framework to address these limitations at both low and high energies \cite{Capozziello:2011et}. Such theories generally incorporate higher-order curvature invariants or non-minimal couplings between scalar fields and geometry in the Einstein--Hilbert action. Notably, these additional terms naturally arise in the one-loop effective action obtained within the semiclassical approach to QFT in curved spacetime \cite{Birrell:book, Adams:1990pn, Cotsakis:2006zn, Amendola:1993bg, Capozziello:2011et, Capozziello:2019klx, DeFelice:2010aj, Clifton:2011jh, Bajardi:2020fxh, Yukawa}.

An alternative approach to addressing the shortcomings of GR is provided by non-local ETGs \cite{Biswas:2005qr, Biswas:2016etb, Biswas:2016egy, Buoninfante:2018xiw, Buoninfante:2018mre}. These theories are generally classified into two main categories: {\it Infinite Derivative Gravity} (IDG) and {\it Integral Kernel Gravity} (IKG). IDG models are constructed from analytic transcendental functions of differential operators, such as exponential functions of the covariant d'Alembert operator $\Box$ \cite{Modesto:2013ioa}. These approaches have been shown to solve classical black hole and Big Bang singularities (see, e.g., Refs.\ \cite{Modesto:2011kw, Briscese:2012ys}). By contrast, IKG models are based on integral kernels of differential operators, most notably the inverse d'Alembert operator $\Box^{-1}$ \cite{Deser:2007jk}. In the latter context, a minimal non-local extension of GR is given by
\begin{equation}
L=\sqrt{-g}\,R\,[1+f(\Box^{-1}R)]\,, 
\end{equation}
where the term $\Box^{-1}R$ can account for the late-time accelerated expansion of the Universe without introducing any dark energy component \cite{capozziello:really,capozziello:weinberg}. From a theoretical perspective, IDG models were originally developed with the aim of constructing fully renormalizable and unitary theories of quantum gravity \cite{Biswas:2011ar}. By contrast, IKG models are motivated by IR quantum corrections arising from QFT in curved spacetime \cite{Barvinsky:2014lja,Modesto:2017sdr,Nojiri:2007uq, Odi1, Odi2,Bahamonde:2017sdo,Bombacigno}).

In general, alternative theories of gravity -- and their corresponding field equations in particular -- are mathematically challenging to handle. It is therefore desirable to introduce a criterion capable of selecting self-consistent models while discarding unviable ones. Dynamical systems governing the cosmic evolution are typically highly nonlinear and, in many cases, integro-differential in nature, making standard analytical techniques difficult to apply.

A powerful method to reduce dynamics and identify exact solutions is provided by the {\it Noether Symmetry Approach} \cite{Capozziello:1996bi, Dialektopoulos:2018qoe}. The presence of Noether symmetries in a given class of models guarantees the existence of conserved quantities, i.e., first integrals of motion, which can render the system integrable. Moreover, the existence of such symmetries is often associated with physically meaningful models \cite{Capozziello:1996bi, Bajardi:2022ypn}. The Noether Symmetry Approach can also be extended to non-local gravity. As discussed in Refs.\ \cite{Capozziello:2012iea, Urban:2020lfk, Bajardi:2019zzs, baj1, Bajardi:2020xfj, Bajardi:2021tul, Capozziello:2012hm, Capozziello:2007wc, Bajardi:2020mdp}, this method provides an effective tool for selecting viable models and deriving exact cosmological solutions.

In this paper, we consider some models of non-local gravity applied to cosmology, whose field equations admit, in a spatially flat Friedmann background, power law solutions $a(t) \sim t^q$ for the scale factor. Such modified cosmologies lead to different thermal histories of the Universe. To see how this happens, the expansion rate $H$ of the Universe can be expressed in terms of the expansion rate $H_{GR}$ of GR as $H(T)\equiv A(T)H_{GR}(T)$, where the factor $A(T)$ encodes the information related to the underlying model of gravity that extends or modifies GR. Typically, the factor $A(T)$ is defined in such a way that the successful predictions of Big Bang Nucleosynthesis (BBN) are preserved, hence $A(T) \neq 1$ at very early times only, i.e.\ in the pre-BBN epoch -- which is a cosmic era not directly constrained by cosmological observations --, and $A(T)\to 1$ when (or just before) the BBN begins. The evolution of the expansion rate of the Universe can be found via Noether symmetries, which, in a cosmological setting, prove to be a powerful tool to integrate the dynamical systems.

The main goal of this paper is to show that non-local generalisations of the Standard Cosmological Model can provide a consistent and unified minimal framework for PeV DM and IceCube/KM3NeT neutrino observations. In fact, the IceCube collaboration \cite{2a,3a}  reported events of neutrinos with energies $\sim1$\,PeV \cite{3a}. Various astrophysical sources are candidates for the generation of these neutrino high-energy events \cite{14a,15a,16a,Sahu:2014fua}, although there is still no clear correlation with known astrophysical hot-spots like supernova remnants or active galactic nuclei \cite{Aartsen:2016oji}. This suggested the possibility that high-energy neutrinos could arise from the decay of PeV-mass DM particles \cite{17a,21a,22a,23a,24a,25a,26a,27a,28a,29a,30a,31a,32a,33a,34a,35a,36a,37a} (see also Refs.\ \cite{Griest:1989wd,Beacom:2006tt,40a,41a,52a,merle,lambIce}).

Here, we explore the possibility to explain the PeV DM relic density as well as the decay rate required for the IceCube/KM3NeT observations with the minimal DM-neutrino four-dimensional interaction $y_{\alpha\chi}\bar L_\alpha H\chi$, where $\alpha$ indicates the mass eigenstates of the three active neutrino families, $\chi$ the DM particle, $H$ the Higgs doublet, $L_\alpha$ the left-handed lepton doublet and $y_{\alpha\chi}$ the Yukawa couplings. In particular, we calculate the freeze-in abundance of DM in non-locally modified cosmologies and show that the couplings required to obtain the correct relic abundance depend on the parameters of the non-local ETGs. By selecting the cosmological parameters so that there is no deviation from the predictions of the BBN, with our minimal model of decaying DM $y_{\alpha\chi}\bar L_\alpha H\chi$, we can satisfy both the relic abundance and the IceCube/KM3NeT neutrinos requirements with a single value of the combination $\sum_\alpha |y_{\alpha\chi}^2|$ of the couplings.

 Let us now briefly review some high-energy processes related to DM detection. The IceCube measurements of ultra-high-energy (UHE) neutrinos at TeV/PeV energies \cite{Aartsen:2013jdh,Aartsen:2014gkd,Aartsen:2017mau} offered novel possibilities for astroparticle physics, such as the astrophysical origin of these neutrinos, which is still not fully understood. The astrophysical processes that could source such highly energetic neutrinos might be ascribed to the decay of charged pions \cite{Loeb:2006tw,Murase:2013rfa,Tamborra:2014xia,Bechtol:2015uqb} or to photo-hadronic interactions \cite{Winter:2013cla,Murase:2015xka}. Moreover, together with neutrinos, gamma rays are also produced, as confirmed by the multi-messenger observation of the gamma ray and neutrino emission from the flaring blazar TXS 0506+056 \cite{IceCube:2018cha,IceCube:2018dnn,Ahnen:2018mvi}. Strong constraints on the contribution of several extragalactic astrophysical sources have been set via spatial and temporal correlation analyses with gamma rays \cite{Adrian-Martinez:2015ver,Aartsen:2018ywr} and angular clustering \cite{Dekker:2018cqu}. The possible astrophysical events are thus genuine cosmic ray accelerators; for these candidates, the gamma ray flux is highly suppressed \cite{Murase:2015xka,Senno:2015tsn}, suggesting that an alternative source for a diffuse UHE neutrino flux could be decaying DM \cite{Chianese:2016opp,Chianese:2016kpu,Chianese:2017nwe,Bhattacharya:2019ucd,Kopp:2015bfa}. It is worth noticing that another available source might be DM annihilation \cite{Bhattacharya:2019ucd,Chianese:2018ijk}, but in general small neutrino fluxes are inferred from the unitarity limit, which turns out not to be detectable. Different DM models are further constrained by means of gamma ray observations. While the current data (neutrinos and gamma rays) show that hadronic final states are strongly disfavoured or excluded, leptonic ones are slightly disfavoured or allowed \cite{Cohen:2016uyg,Blanco:2018esa}. The most favourable case is that of a heavy DM candidate coupled only to neutrinos. Observations related to gamma ray constraints, such as those from the Fermi-LAT experiment, have excluded or disfavoured various DM decay channels. In this sense, it is worth exploring whether gamma ray experiments in different energy ranges will allow us to place further constraints on the existence of decaying DM fluxes. The lower energy range ($\lesssim10^2$\,GeV) is mainly explored by Fermi-LAT \cite{Atwood:2009ez}, while the higher energy range ($\gtrsim 10^2$\,TeV) is explored by air shower experiments \cite{Kalashev:2016cre}. These experiments are used to constrain annihilating or decaying DM through data by Pierre Auger \cite{Aab:2015bza}, CASA-MIA \cite{Chantell:1997gs}, KASCADE \cite{Kang:2015gpa} and LHAASO \cite{Cao:2010zz}. Finally, we mention the Alpha Magnetic Spectrometer (AMS-02) too, which has produced high-precision measurements of the local cosmic ray antiproton spectrum. This is considered to be a critical channel for indirect DM searches \cite{[1],[2]}. In Refs.\ \cite{[3],[4]}, it was found that there is statistically significant evidence for an excess over the expected astrophysical background in the cosmic ray antiproton spectrum (with energies $\sim10$\,GeV). Their best-fit suggested the annihilation of DM particles with a mass of $\sim60$\,GeV. Compared to the cosmic ray positron excess \cite{[5],[6],[7]}, observed by PAMELA \cite{[8]} and later by AMS-02 \cite{[9]}, the DM interpretation of the antiproton excess is consistent with a standard thermal WIMP.

This paper is organized as follows. In Sec.\ 2, we review the main features of non-local cosmological models, formulated as dynamical systems governing cosmic evolution, and summarise the corresponding solutions obtained through the Noether Symmetry Approach. In Sec.\ 3, we show how the observed DM relic abundance and the IceCube/KM3NeT neutrino data can be consistently accommodated within a minimal DM framework, assuming a cosmological background described by non-local ETGs. In particular, we consider models containing $\Box^{-1}$ contributions to geometric invariants such as the Ricci scalar $R$ and the torsion scalar $T$. The underlying idea is that non-local terms of the form $\Box^{-1}R$ or $\Box^{-1}T$ can be recast in terms of auxiliary scalar fields through suitable redefinitions like $R=\Box \phi$ and $T=\Box \varphi$. A similar procedure applies to boundary terms such as $B=\nabla^{\mu}T_{\mu}$. Within this formulation, non-local gravity models can be treated within the standard of scalar-tensor theories of gravity and then as Horndeski theories. Sec.\ 4 is devoted to a discussion of the results and to the conclusions.

\section{Dynamical Systems from Non-local Gravity Cosmology}
\setcounter{equation}{0}

In this section we solve dynamical systems coming from  non-local gravity cosmology. Exact solutions are inferred from the Noether Symmetry Approach, which we briefly illustrate below.

According to the previous considerations, non-local gravity contributions in the field equations implies the presence of characteristic lengths and masses \cite{Modesto:2017sdr}.
As a general case, let us take into account a non-local theory of gravity that is dynamically described by curvature. The field equations can be written in the form
\begin{equation}
    G_{\mu\nu}+\Xi_{\mu\nu}^\textup{non-local}=\kappa T_{\mu\nu},
\end{equation}
where $G_{\mu\nu}$ is the Einstein curvature tensor. $\Xi_{\mu\nu}^\textup{non-local}$ refers to the non-local curvature part, potentially containing higher-order terms as well as powers of the $\Box^{-1}$ operator in the curvature invariants \cite{Capozziello:2021krv}. $T_{\mu\nu}$ is the energy-momentum tensor of matter. The field equations can be recast in the form
\begin{equation}
    G_{\mu\nu}=\kappa T^\textup{eff}_{\mu\nu},\qquad\text{where}\qquad T^\textup{eff}_{\mu\nu}\equiv T_{\mu\nu}-\frac{1}{\kappa}\Xi_{\mu\nu}^\textup{non-local},
\end{equation}
so that the non-local terms can be interpreted as an \emph{effective fluid} sourcing the Einstein field equations. According to this interpretation, we can assume $T_{\mu\nu}\ll \frac{1}{\kappa}\Xi_{\mu\nu}^\textup{non-local}$, that is although ordinary matter is present (as we are mainly interested in the radiation-dominated epoch), the non-local curvature terms are dominant.

This means that the aim is to search, via the Noether symmetry approach, for solutions of the field equations in vacuum. In particular, we shall focus on power law solutions of the form $a=a_*(t/t_*)^q$, fixing $t_*$ to the time (corresponding to the temperature $T_*$) at which the Universe starts to evolve according to the standard cosmological model, so that then $q=1/2$. To avoid inconsistencies with the predictions of the BBN, we restrict the dynamical system analysis to the pre-BBN era, thus $T_* \gg T_\textup{BBN}\sim 1$\,MeV.

We assume that the Universe is described by a spatially flat Friedmann--Robertson--Walker metric
\begin{equation}\label{FRW}
  ds^2 = dt^2-a^2(t)(dx^2+dy^2+dz^2)\,,
\end{equation}
where $a(t)$ is the scale factor. The energy density of relativistic matter (radiation) is given by $\rho=\pi^2 g_* T^4/30$, with $g_*=106$, while the Hubble rate is defined as $H\equiv\dot a/a$. From the conservation of entropy, one finds that $T(t) a(t) = a_0 T_0 = \textup{const.}$, where $a_0=1$ and $T_0 \sim 10^{-4}$\,eV at the present time. In the following of this section, we will specialise to some specific non-local gravity models derived from the Noether Symmetry Approach. From a physical point of view, the goal is to test whether the recent UHE neutrino signals can be a signature of non-local gravity in the early Universe.

\subsection{The Noether Symmetry Approach to reduce dynamical systems}
Before taking into account particular non-local gravity models, let us remind some basic concepts related to the Noether Symmetry Approach used to reduce, and eventually integrate, dynamical systems. A detailed exposition of the method can be found in Refs.\ \cite{Capozziello:1996bi, Bajardi:2022ypn}. In general terms, the Noether symmetries of a Lagrangian are useful to reduce the dynamics and analytically solve systems of differential equations. See also Ref.\ \cite{SalvConstraining}.

Let us consider a point transformation $(x,y) \to (\bar{x},\bar{y})$ and let $\varepsilon$ be an arbitrary real parameter such that
\begin{equation}
\label{nsa1.2}
\bar{x}=\bar{x}(x,y;\varepsilon),  \qquad \bar{y}=\bar{y}(x,y;\varepsilon).
\end{equation}
The first-order Taylor expansion of the infinitesimal transformation \eqref{nsa1.2} around $\varepsilon=0$ yields
\begin{align}
&\bar{x}(x,y;\varepsilon)=x+\varepsilon\frac{\partial\bar{x}}{\partial\varepsilon}\bigg|_{\varepsilon=0}  %
                                       =x+\varepsilon\xi(x,y)   \label{nsa1.3}\,, \\
&\bar{y}(x,y;\varepsilon)=y+\varepsilon\frac{\partial\bar{y}}{\partial\varepsilon}\bigg|_{\varepsilon=0}  %
                                       =y+\varepsilon\eta(x,y)  \label{nsa1.4} \,.
\end{align}
The functions $\xi(x,y)$ and $\eta(x,y)$ are the components of the tangent vector $\mathbf{X}$ to the orbit of the transformation at the point $(x,y)$, i.e.\
\begin{equation}
\label{nsa1.5}
\mathbf{X}=\xi(x,y)\frac{\partial}{\partial x}+\eta(x,y)\frac{\partial}{\partial y}\,.
\end{equation}
Since $(x,y)$ is an arbitrary point, Eq.\ \eqref{nsa1.5} provides the tangent vector field to the group orbits, the so-called infinitesimal generator of the one-parameter group of point transformations. The prolongation of the tangent vector, involving the $n$-th derivatives, can be computed by means of the following relations:

\begin{align}
&\bar{y}' \equiv \frac{d\bar{y}(x,y;\varepsilon)}{d\bar{x}(x,y;\varepsilon)}%
                 =\frac{y'(\partial\bar{y}/\partial y)+(\partial\bar{y}/\partial x)}{y'(\partial\bar{x}/\partial y)+(\partial\bar{x}/\partial x)}%
                 =\bar{y}'(x,y,y';\varepsilon),    \label{nsa1.6}  \\
&\bar{y}''\equiv \frac{d\bar{y}'}{d\bar{x}}=\bar{y}''(x,y,y',y'';\varepsilon), \label{nsa1.7} \\
&\ldots \notag
\end{align}
The prolongations of the generator $\mathbf{X}$ can be obtained through a first-order Taylor expansion around $\varepsilon=0$. Replacing Eqs.\ \eqref{nsa1.3} and \eqref{nsa1.4} into Eqs.\ \eqref{nsa1.6} and \eqref{nsa1.7}, the $n$-th derivatives of the transformed coordinates (up to the first order) read
\begin{align}
&\bar{y}'=y'+\varepsilon\left(\frac{d\eta}{dx}-y'\frac{d\xi}{dx}\right) =y'+\varepsilon\eta^{[1]}  \,,   \\
&\vdots   \notag\\
&\bar{y}^{(n)}=y^{(n)}+\varepsilon\left(\frac{d\eta^{(n-1)}}{dx}-y^{(n)}\frac{d\xi}{dx}\right)  %
                         =y^{(n)}+\varepsilon\eta^{[n]}  \,,
\end{align}
where
\begin{equation}
\label{nsa1.10}
\eta^{[n]} \equiv \frac{d\eta^{(n-1)}}{dx}-y^{(n)}\frac{d\xi}{dx}=\frac{d^n}{dx^n}(\eta-y'\xi)+y^{(n+1)}\xi \,
\end{equation}
is the $n$-th prolongation function of $\eta$. Therefore, the $n$-th prolongation of Noether's vector can be written as
\begin{equation}
\label{nsa1.12}
\mathbf{X}^{[n]}=\mathbf{X}+\eta^{[1]}\partial_{y'} +...+\eta^{[n]}\partial_{y^{(n)}}\,.
\end{equation}
Starting from Eq.\ \eqref{nsa1.12} and considering a point-like Lagrangian $L=L\left(t,q(t),\dot{q}(t)\right)$, with $q^i$ being the coordinates and $t$ the time, the first prolongation of Noether's vector can be expressed as
\begin{equation}
\label{nsa1.26}
\mathbf{X}^{[1]}=\xi(t,q)\frac{\partial}{\partial t}+\eta^i(t,q)\frac{\partial}{\partial q^i}%
                                +(\dot{\eta}^i-\dot{\xi}\dot{q}^i) \frac{\partial}{\partial\dot{q}^{i}} \,.
\end{equation}
The first Noether theorem states that the one-parameter group of point transformations generated by $\mathbf{X}$ is a one-parameter group of Noether point symmetries for the dynamical system described by $L$ if and only if there exists a function $g\bigl(t,q(t)\bigr)$ such that
\begin{equation}
\label{nsa1.27}
\mathbf{X}^{[1]}L+\dot{\xi}L=\dot{g}\,,
\end{equation}
 whose associated first integral of motion is
\begin{equation}
\label{nsa1.28}
I(t,q,\dot{q})=\xi \left(\dot{q}^i \frac{\partial L}{\partial\dot{q}^i}-L \right)-\eta^i \frac{\partial L}{\partial\dot{q}^i}+g\,.
\end{equation}
This integral allows to ``reduce'' the degrees of freedom of the dynamical system. If the number of first integrals of motion coincides with the dimension of the configuration space, then the system is completely integrable. In several applications, it is useful to consider  the restricted Noether condition \cite{Capozziello:1996bi}
\begin{equation}
\mathbf{X}L=0\,,
\end{equation}
related to the Noether vector
\begin{equation}
\mathbf{X}=\eta^i(q)\frac{\partial}{\partial q^i}+\dot{\eta}^i \frac{\partial}{\partial\dot{q}^{i}} \,.
\end{equation}
This gives rise to the integral of motion
\begin{equation}
I(q,\dot{q})=\eta^i \frac{\partial L}{\partial\dot{q}^i}\,,
\end{equation}
which is particularly useful for dealing with the internal variables of a dynamical system \cite{Bajardi:2022ypn}. In any case, after selecting the integrals of motion and reducing dynamics, the method allows us: $i)$ to identify the cyclic variables; $ii)$ to fix the form of couplings and potentials in the Lagrangian; $iii)$ to solve exactly the dynamics. Below, we will apply the approach to some significant non-local gravity cosmological models.

\subsection{Curvature non-local gravity }

A possible action for a non-local gravity, based on curvature, is \cite{acunzo}
\begin{equation}
\label{NL:22}
\mathcal{S}=\int\!d^4x \sqrt{-g}\,F\bigl(R,\square^{-1}R\bigr)\,.
\end{equation}
This effective theory is a generalisation of $F(R)$-gravity  including non-local terms, as a consequence of the operators $\Box^{-1}R$. A point-like Lagrangian that is useful for cosmological considerations can be derived after the ``scalarisation'' and the ``localisation'' of the $\Box^{-1}$ operator. Using an auxiliary scalar field $\phi$, we have
\begin{equation}
\phi\equiv\Box^{-1}R\qquad\text{or}\qquad R\equiv\Box\phi.
\end{equation}
In the cosmological metric \eqref{FRW}, the above quantities are given explicitly by
\begin{equation}
    R=-6\biggl(\frac{\ddot a}{a}+H^2\biggr)\,,\qquad \Box\phi=\ddot\phi+3H\dot\phi.
\end{equation}
In such a way, the action can be rewritten in a \emph{local} form and the initial theory can be recast in terms of a scalar-tensor theory described as
\begin{equation}
S=\int d^4x\,\sqrt{-g}\,F(R,\phi)\,,
\end{equation}
and then further improved with the introduction of a Lagrange multiplier $\lambda$. Therefore, after integrating out the second derivatives from the non-local action \eqref{NL:22} specified for the cosmological metric \eqref{FRW}, in the configuration space, that is the minisuperspace $\{a,R,\phi,\lambda\}$, the cosmological point-like Lagrangian is \cite{acunzo}
\begin{equation}
\label{pointlike1}
L=a^3F-a^3\dot\phi\dot\lambda-a^3R\partial_RF+6a\dot a^2\partial_RF-6a\dot a^2\lambda+6a^2\dot a\dot R\partial_{RR}F+6a^2\dot a\dot\phi\partial_{R\phi}F-6a^2\dot a\dot\lambda.
\end{equation}
By deriving the Euler--Lagrange equations from the Lagrangian \eqref{pointlike1} and adopting the above Noether Symmetry Approach (see \cite{acunzo} for details), the following exact  solutions are found:
\begin{eqnarray}
&a(t)&=a_0\,t^p,  \\
&R(t)&=-6\left(-\frac{p}{t^2}+\frac{2 p^2}{t^2}\right),     \\
&\phi(t)&=C_2-\frac{6 p (2 p-1) \log(t)}{3 p-1}\,, \\
&\lambda(t)&=\frac{6^n (3 p-1)\,ks\, e^{C_2 s} \left(\frac{(1-2 p)p}{t^2}\right)^n t^{2-\frac{6 p (2 p-1) s}{3   p-1}}}%
                   {\left(-2 n-\frac{6 p (2 p-1) s}{3 p-1}+2\right)\left(n (6 p-2)+3 p^2 (4 s-3)-6 p s+1\right)}+%
                    \frac{C_3 \, t^{1-3 p}}{1-3 p} - \tilde{c}_1 \,,\\
& s = &\frac{3p-1}{3(2p-1)}\neq 0.
\end{eqnarray}
The Noether symmetry selects also the form of the Lagrangian, which is
\begin{equation}
F(R, \phi) =-\tilde{c}_1 R + kR\, e^{\frac{3p-1}{3(2p-1)}\phi}\,.
\end{equation}
The parameters $n$, $p$ and $s$ are related to each other as follows \cite{acunzo}:
\begin{equation}\label{npsRelation1}
\begin{split}
 &4 n^4 (3 p-1)^3+2 n^3 (1-3 p)^2 \bigl[6 p^2 (6 s-1)-p (18 s+13)+5\bigr]\\
 &\,\,+n^2 (3 p-1) \Bigl[9 p^4  (48 s^2-16 s-3)-3 p^3  (144 s^2+80 s-33)\\
 &\,\,+3 p^2  (36 s^2+92 s+5)-3 p (20 s+13)+8\Bigr]+n\Bigl[54 p^6 s (16 s^2-8 s-3)\\
 &\,\,-9 p^5  (144 s^3+56 s^2-39 s-27)+9 p^4  (72 s^3+132 s^2+31 s-51)\\
 &\,\,-9 p^3  (12s^3+66 s^2+61 s-27)+3 p^2  (30 s^2+77 s-9t)-10 p (3 s+1)+2\Bigr]\\
 &\,\,-3 (1-3 p)^2 p (2 p-1)\bigl[3 p^2 s (4 s-3)-p  (6 s^2+s-3)+2 s-1\bigr]=0,
\end{split}
\end{equation}

\begin{equation}\label{npsRelation2}
 \begin{split}
  &2 n^3 (1-3 p)^2+n^2 (3 p-1) \bigl[3 p^2 (8 s+1)-4 p (3 s+4)+5\bigr]\\
  &\,\,+n \Bigl[18 p^4 s (4 s+1)-3 p^3 (24 s^2+35 s+9)+6 p^2 (3 s^2+13 s+9)-3 p (5 s+9)+4\Bigr]\\
  &\,\,-(1-3 p)^2 (2 p-1) (3 p s-1)=0,
 \end{split}
 \end{equation}
where $s$ is a constant which appears in the solution for the Lagrange multiplier $\lambda$ (see Refs.\ \cite{SalvConstraining, acunzo}). It is worth stressing again that such relations are fixed by the presence of the Noether symmetry, which allows to solve the dynamical system.

\subsection{Teleparallel non-local gravity }
As discussed in the Introduction, the gravitational interaction can be described also by other geometrical invariants derived from torsion and non-metricity \cite{Carmen,Sara}. Here, we want to show that the Noether Symmetry Approach works also for dynamical systems derived from teleparallel gravity.

By including a non-vanishing anti-symmetric part in the affine connection, it is possible to define a non-vanishing rank-3 tensor $T^\alpha_{\mu \nu} \equiv \Gamma^\alpha_{\mu \nu} - \Gamma^\alpha_{\nu \mu}$. In this formalism, torsion thus appears as the fundamental gravitational field, at the same level as curvature, giving rise to additional degrees of freedom. The rank-3 tensor $T^\alpha_{\mu \nu}$ is the \textit{torsion tensor}, and a specific contraction of its indices allows to define the \textit{torsion scalar} $T$. The gravitational theory describing  spacetime dynamics only by torsion and without curvature is the so-called Teleparallel Equivalent of General Relativity (TEGR) \cite{Ferraro2,Linder:2010py,Pereira.book,Cai18,Nesseris:2013jea,Ferraro3,TJCAP16}. The teleparallel Lagrangian differs from the Einstein--Hilbert one only because of a four-divergence, which means that TEGR and GR are dynamically equivalent \cite{Carmen, Hammond:2002rm, Maluf:2013gaa, Aldrovandi:2015wfa, Aldrovandi:2013wha,Cai:2015emx, Li:2010cg, Wu:2010mn, Capozziello:2011hj, Boehmer:2011gw, Wu:2010xk, Iorio:2012cm, Bahamonde:2019zea, Bahamonde:2016grb}. In Refs.\ \cite{Aldrovandi:2015wfa, Aldrovandi:2013wha}, it was shown that TEGR can be recast as a gauge theory of the translation group.

The previous model can be considered also in a non-local teleparallel framework \cite{SalvConstraining}, assuming
\begin{equation}\label{NLT:22}
\mathcal{S}=\int\!d^4x \sqrt{-g}\,F\bigl(T,\square^{-1}T\bigr)= \int\!d^4x \sqrt{-g}\, Tf(\square^{-1}T)\,.
\end{equation}
In this case, the torsion scalar $T$ is coupled to a non-local function of $T$ itself, that is one can write, analogously as in the above curvature case, $\square^{-1}T=\phi$. Searching for Noether symmetries, the Lagrangian in Eq.\ \eqref{NLT:22} can be reduced to \cite{SalvConstraining}
\begin{equation}\label{eq:teleparallel1}
L=-6a\dot a^2\Bigl(c_2e^{\frac{c_3\phi}{c_1}}-\theta-1\Bigr)-a^3\dot\theta\dot\phi\,,
\end{equation}
and thus to the solutions
\begin{eqnarray}
a(t) &=& a_0t^p\,, \quad  \phi(t)=\frac{6 p^2 \log (t-3 p t)}{1-3 p}\,,\\
\theta(t) &=& c_2 (1-3p)^{3-3p} t^{2-3p}+\frac{\theta_0 t^{1-3 p}}{1-3 p}-1\,, \\
 f(\phi) &=& c_2e^{\frac{\left(9 p^2-9 p+2\right) \phi }{6 p^2}}\,,\label{eq:teleparallel4}
\end{eqnarray}
with $p\neq 1/3$. Here $\theta$ is the Lagrange multiplier defined according to the relation $\theta(\Box \phi-T)$ to be inserted in the action \eqref{NLT:22} for consistency (see also below). In Ref.\ \cite{SalvConstraining}, all the details of this procedure are reported. Here we are interested in the related dynamical system and its exact solutions.

\subsection{Teleparallel non-local gravity including a boundary term}
As shown in Ref.\ \cite{Sara}, it is possible to restore the full equivalence between the curvature and the torsion representations of gravity by including boundary terms into the gravitational dynamics. Specifically, we have
\begin{equation}
R= -T +\frac{2}{e}\partial_{\mu}(eT^{\mu})= -T + B\,,
\end{equation}
where $B$ is the boundary term \cite{SalvConstraining}. See also Ref.\ \cite{Capriolo} for a discussion.

Let us then consider the following generalised non-local action involving non-local contributions from both the torsion scalar $T$ and the boundary term $B$. We have
\begin{equation}\label{general-non-local}
	\mathcal{S}= -\frac{1}{2\kappa}\int d^{4}x\, e T +  \frac{1}{2\kappa}\int d^{4}x\, e\left[ \, (\xi T+{\tilde \chi} B)f(\phi,\varphi) +\theta (\square \phi - T)+  \zeta (\square \varphi -B)\right]+ \int d^{4}x\, e\,L_{m} \,.
\end{equation}
The Lagrange multipliers $\theta$ and $\zeta$ constrain the scalar fields to be
\begin{equation}
\phi=\Box^{-1}T,\qquad\varphi=\Box^{-1}B.
\end{equation}
The standard Ricci scalar is recovered in the action \eqref{general-non-local} for the choice $\xi=-{\tilde \chi}=-1$ of the coupling constants.


In the cosmological background \eqref{FRW}, the two field equations governing the evolution of this non-local dynamical system are
\begin{eqnarray}
    3H^2(\theta-\xi f+1)&=&\frac{1}{2}\dot\zeta\dot\varphi+\frac{1}{2}\dot\theta\dot\phi+3H(\dot\zeta-{\tilde \chi}\dot f)+\kappa\rho\,,\\
    (2\dot H+3H^2)(\theta-\xi f+1)&=&-\frac{1}{2}\dot\zeta\dot\varphi-\frac{1}{2}\dot\theta\dot\phi-\dot f[2H(\xi+2{\tilde \chi})+{\tilde \chi}]+2H(2\dot\zeta+\dot\theta)+\ddot\zeta-\kappa p\,,
\end{eqnarray}
where $\kappa=8\pi G$. Here $\rho$ and $p$ are respectively the energy density and the pressure of an ideal cosmic fluid. The  cosmological point-like Lagrangian is now \cite{SalvConstraining}
\begin{equation}
L=6\chi a^2\dot a\dot\phi f_\phi+6\chi a^2\dot a\dot\varphi f_\varphi-6\xi a\dot a^2f-6a^2\dot a\dot\zeta+6a\theta\dot a^2+6a\dot a^2-a^3\dot\zeta\dot\varphi-a^3\dot\theta\dot\phi\,,
\end{equation}
and the corresponding cosmological solutions, obtained from the Noether symmetries, are
\begin{eqnarray}
a(t) &=& t^p\,,\\ \phi (t)&=&\frac{6 p^2 \log (t-3 p t)}{1-3 p}\,,  \\
\varphi (t)&=& -6 p \log t\,, \\
\theta (t)&=&\frac{6 c_{11} c_7 p t^{2-3 p} (1-3 p)^{\frac{c_7 (2-3 p) p}{-3 c_5 p+c_5+c_7 p}} (p (\xi +3 \chi )-\chi )}{3 p-2}+\frac{\theta_{1} t^{1-3 p}}{1-3 p}\,, \\
\zeta (t) &=& -\frac{6 c_{11} c_5 p t^{2-3 p} (1-3 p)^{-\frac{c_7 p (3 p-2)}{-3 c_5 p+c_5+c_7 p}} (p (\xi +3 \chi )-\chi )}{3 p-2}+\zeta_{0}+\frac{\zeta_{1} t^{1-3 p}}{1-3 p}\,, \\
f(\phi,\varphi) &=&  \frac{1}{\xi }-\frac{6 c_{11} p (-3 c_5 p+c_5+c_7 p) }{9 p^2-9 p+2}\exp \left(-\frac{\left(9 p^2-9 p+2\right) (c_5 \varphi -c_7 \phi )}{6 p (-3 c_5 p+c_5+c_7 p)}\right)\,,
\end{eqnarray}
with $p\neq 0, 1/3, 2/3$. It is worth noticing that the form of $f(\phi,\varphi)$ in the Lagrangian is fixed by the presence of the Noether symmetries. As shown in Ref.\ \cite{SalvConstraining}, this cosmological model is consistent only for $p<1/3$.

\subsection{Horndeski Gravity}
In all of the above models, a prominent role is played by the scalarisation process. According to this picture, non-local  theories of gravity, in particular IKG theories, can be reduced to scalar-tensor models. With this consideration in mind, we have to point out that Horndeski has taken into account the most complete scalar-tensor theory of gravity with second-order derivatives in the action but still with second-order equations of motion \cite{horndeski}. The action is given by the sum of the following Lagrangian densities:
\begin{equation}\label{eq1}
\mathcal{S}_\textup{Horndeski} = \sum _{i=2} ^5 \int d^4 x \sqrt{-g}\mathcal{L}_i\,,
\end{equation}
where
\begin{eqnarray}\label{eq2}
\mathcal{L}_2 &=& G_2\left(\phi , X\right) \,, \\
\mathcal{L}_3 &=& G_3\left(\phi , X\right)\square \phi \,, \\
\mathcal{L}_4 &=& G_4\left(\phi , X\right)R + G_{4,X}(\phi,X) \left[\left( \square \phi\right)^2 -\left( \nabla _{\mu}\nabla _{\nu}\phi\right)^2\right] \,, \\
\mathcal{L}_5 &=& G_5\left(\phi , X\right) G_{\mu\nu} \nabla ^{\mu}\nabla ^{\nu} \phi - \frac{1}{6}G_{5,X}(\phi,X)\left[\left(\square \phi\right)^3- 3\square \phi \left( \nabla _{\mu}\nabla _{\nu}\phi\right)^2 + 2 \left( \nabla _{\mu}\nabla _{\nu}\phi\right)^3\right]\,.\label{eq22}
\end{eqnarray}
The functions $G_2\left(\phi, X\right)$, $G_3 \left(\phi, X\right)$, $G_4\left(\phi, X\right)$ and $G_5\left(\phi, X\right)$ are arbitrary functions of the scalar field $\phi$ and the kinetic term $X\equiv- \frac{1}{2}\left(\nabla \phi\right)^2 =-\frac{1}{2} \nabla^{\mu}\phi \nabla_{\mu}\phi$. The complete form of the field equations is rather convoluted. It can be found in  Ref.\ \cite{SalvH}. Here we will focus on the Noether Symmetry Approach for the case $G_3(\phi,X)=G_3(\phi)$. The cosmological point-like Lagrangian in the minisuperspace $\{a,\phi\}$ is then
\begin{equation}
L=a^3G_2-a^3G_3'\dot\phi^2-6aG_4\dot a^2-6a^2G_{4\phi}\dot a\dot\phi+3a(2G_{4X}-G_{5\phi})\dot a^2\dot\phi^2+G_{5X}\dot a^3\dot\phi^3\,.
\end{equation}
From the Noether symmetry, it is possible to obtain the general solutions
\begin{equation}\label{Horneskiscale}
a(t) = \alpha_0   t ^{(\alpha_1+2\xi_1)/3\xi_1},\qquad\phi (t) = \phi_0 t - \frac{\phi_1}{\xi_1}\,,
\end{equation}
where the parameters $\alpha_1$, $\xi_1$ and $\phi_{0,1}$ are  integration constants. In Ref.\ \cite{SalvH}, a general discussion on Horndeski models selected by the Noether symmetries is reported. For the aim of this paper, it is important to say that non-local gravity can give rise to cosmological models  which can be exactly solved. We want to show that different dynamical systems representing the evolution of the Universe can be highly sensitive to the parameters of non-local gravity at energies probed by the IceCube/KM3NeT collaborations. In the following, we are going to discuss this issue considering the most recent releases about high-energy neutrinos.

\section{PeV neutrinos in non-local cosmologies}
\setcounter{equation}{0}

We proceed with our analysis by recalling the main features related to the DM relic abundance and the IceCube data in modified cosmologies. As discussed in the previous section, the Noether Symmetry Approach reveals that, despite being \textit{a priori} very different dynamical systems for the gravitational interaction, several non-local cosmological models admit power law solutions of the form
 \begin{equation}\label{powerlaw}
   a(t) = a_0  \left(\frac{t}{t_0}\right)^q\,.
 \end{equation}
Our next aim is then to express the expansion rate of the Universe in terms of the temperature $T$ according to the parametrisation \cite{BD}
 \begin{equation}\label{H=AHGRIce}
   H(T)=A(T)H_{GR}(T)\,,
 \end{equation}
where $H_{GR}$ is the expansion rate in General Relativity ($H_{GR}=\sqrt{\frac{\rho}{3M_P^2}}=\sqrt{\frac{\pi^2 g_*}{90}}\, \frac{T^2}{M_P}$), while $A(T)$ is the so-called amplification factor. Using the power law solution \eqref{powerlaw} and the relation $Ta=T_0$ (as usual, we assume an adiabatic expansion of the Universe and $T_0\simeq 10^{-4}$\,eV is its present temperature), the amplification factor can be rewritten in the form
 \begin{eqnarray}\label{A(T)Ice}
   A(T) &=&\eta \left(\frac{T}{T_*}\right)^\nu\,, \\
   \nu &=& \frac{1}{q}-2\,, \qquad \eta = 1\,, \nonumber
 \end{eqnarray}
where the subscript ``$*$'' indicates the temperature $T_*$ (or, equivalently, the time $t_*$) at which the Universe starts to evolve according to the Standard Cosmological Model, while $\eta$ and $\nu$ are the parameters characterising the cosmological model under consideration.\footnote{It is worth noticing that the parametrisation \eqref{A(T)Ice} occurs in different {\it modified} cosmological scenarios \cite{fornengoST,BD,gondolo}, with the parameter $\nu$ being a useful label for discerning the various dynamical systems: $\nu=2$ in Randall--Sundrum type II brane cosmology \cite{randal}, $\nu=1$ in kination models \cite{kination}, $\nu=0$ in cosmologies with an overall boost of the Hubble expansion rate \cite{fornengoST}, $\nu=-0.8$ in scalar-tensor cosmology \cite{fornengoST}, $\nu=2/n-2$ in $f(R)$ cosmology with $f(R)=R+\alpha R^n$ \cite{lamb}.} To preserve the successful predictions of the BBN, hereafter we refer to the pre-BBN epoch, which is a phase of the cosmic evolution not directly constrained by cosmological observations.   Therefore, we require that $A(T) \neq 1$ at early times (for $T\gtrsim T_* \gg T_\textup{BBN}$) and $A(T)\rightarrow1$ for $T=T_*$ (before the BBN begins), that is
\[
A(T) = 
\left\{ \begin{array}{ll}  \eta \left(\frac{T}{T_*}\right)^\nu & \quad T\gtrsim T_* \gg T_\textup{BBN} \\
       1 & \quad T\lesssim T_{*} \end{array} \right.
\]
As we will discuss below, the transition temperature is fixed at $T_* \gtrsim m_\chi$.\footnote{ Here we are following Ref.\ \cite{BD}, in which the enhancement function $A(T)$ is parametrised as
 \begin{equation}\label{A(T)2}
    A(T)=\left\{ \begin{array}{lcr}
    1+\eta\left(\frac{T}{T_*}\right)^\nu \tanh \frac{T-T_{re}}{T_{re}} & \mbox{for} & T>T_\textup{BBN} \\
    1 & \mbox{for} & T\leq T_\textup{BBN} \end{array} \right.\,.
 \end{equation}
Eq.\ \eqref{A(T)2} is a suitable reparametrisation to describe a modified cosmology that {\it smoothly} becomes the standard cosmology at some {\it re-entering} temperature $T_{re}$. The latter is conventionally fixed at the BBN temperature, $T_{re}\sim 1$\,MeV, a condition that avoids conflicts with BBN predictions. In the regime $T\gg T_\textup{BBN}$, the function $A(T)$ behaves as $A(T)\simeq \eta\left(\frac{T}{T_f}\right)^\nu$.}

To explain the IceCube high-energy signal in the framework of the cosmological backgrounds described by non-local cosmologies, we consider the four-dimensional operator \cite{merle}
 \begin{equation}\label{4dop}
 {\cal L}_{d=4}=y_{\alpha\chi}\bar L_\alpha H\chi\,, \qquad \alpha = e, \mu, \tau\,,
 \end{equation}
where $\chi$ is the DM particle, $H$ the Higgs doublet, $L_\alpha$ the left-handed lepton doublet corresponding to the generation $\alpha$ ($=e,\mu,\tau$) and $y_{\alpha \chi}$ the Yukawa couplings. We restrict the analysis to the freeze-in production, which means that DM particles are never in thermal equilibrium owing to their very weak interactions, so that they are gradually produced from the hot thermal bath \cite{hall,merle}. This is possible because of the feeble coupling to the standard matter particles for temperatures $T\gg m_\chi$, allowing, in such a way, DM particles to remain in the Universe due to the smallness of the back-reaction rates and the slowness of their decay rate. A sizeable DM abundance is hence allowed to form in this scenario until temperatures of the order $T\sim m_\chi$ are reached (for temperatures below $m_\chi$, instead, the phase space of the DM particles is not allowed kinematically). The DM particle evolution is governed by the Boltzmann equation. Denoting by $Y_\chi=n_\chi/s$ the DM abundance, where $n_\chi$ is the number density of the DM particles and $s=\frac{2\pi^2}{45}g_*(T)T^3$ the entropy density (with $g_*\sim 106.7$ being the degrees of freedom of relativistic particles), from the Boltzmann equation one obtains
 \begin{equation}\label{Boltz}
   \frac{dY_\chi}{dT}=-\frac{1}{H T s}\left[\frac{g_\chi}{(2\pi)^3}\int C\frac{d^3p_\chi}{E_\chi}\right]\,,
 \end{equation}
where $H$ is the expansion rate of the Universe in a modified cosmology and $C$ the collision term. The DM relic abundance reads
 \begin{equation}\label{DM1}
   \Omega_\textup{DM}h^2=\frac{2 m_\chi^2 s_0 h^2}{\rho_c}
   \int_0^\infty\frac{dx}{x^2}\left(-\frac{dY_\chi}{dT}\Big|_{T=\frac{m_\chi}{x}}\right)\,,
 \end{equation}
where $x=m_\chi/T$, $s_0=\frac{2\pi^2}{45}g_*T_0^3\simeq 2891.2/$cm$^3$ is the present value of the entropy density and $\rho_c=1.054\cdot 10^{-5}h^2$\,GeV/cm$^3$ is the critical density of the Universe \cite{Planck}.

The dominant contributions to the DM production, corresponding to the four-dimensional operator \eqref{4dop}, are essentially two: $1)$ the {\it inverse decay} processes $\nu_\alpha+H^0\to \chi$ and $l_\alpha + H^+\to \chi$, occurring when $m_\chi> m_H+m_{\nu, l}$ and proportional to the factor $|y_{\alpha \chi}|^2$; $2)$ the {\it Yukawa production} processes such as $t+{\bar t}\to {\bar \nu}_\alpha+\chi$, which is proportional to $|y_{\alpha \chi} y_t|^2$ ($t$ being the top quark here). Therefore, one has that
 \begin{eqnarray}\label{dYtot}
 \frac{dY_\chi}{dT} &=& \frac{dY_\chi}{dT}\Big|_{inv.dec.}+\frac{dY_\chi}{dT}\Big|_{Yuk.prod.} \\
  &\simeq & \frac{dY_\chi}{dT}\Big|_{inv.dec.} = -\frac{m_\chi^2 \Gamma_\chi}{\pi^2 H s}\, K_1\left(\frac{m_\chi}{T}\right)\,, \label{dYinvdec}\,
 \end{eqnarray}
where we used the fact that the term $\frac{dY_\chi}{dT}\Big|_{inv.dec.}$ is dominant with respect to $\frac{dY_\chi}{dT}\Big|_{Yuk.prod.}$ (see Appendix A).
%
%
In the last equation, the interaction rate $\Gamma_\chi$ is given by
 \begin{equation}\label{ratechi}
   \Gamma_\chi=\sum_\alpha \frac{|y_{\alpha\chi}|^2}{8\pi}\, m_\chi\,, \qquad \alpha=e, \mu, \tau\,,
 \end{equation}
with $K_1(x)$ being a modified Bessel function of the second kind. Therefore, one finally finds
\begin{equation}\label{Heta}
 H s =\frac{3.32\pi^2}{45}g_*^{3/2}\eta\left(\frac{m_\chi}{T_*}\right)^\nu \frac{1}{{\tilde x}^{5-\nu}}\,,
  \qquad {\tilde x}\equiv \frac{T}{T_*}\,.
\end{equation}
The relic abundance determined by the inverse decay term is then
 \begin{eqnarray}
   \nonumber\Omega_\textup{DM}h^2 &=& \frac{45h^2}{1.66\pi^2 g^{3/2}}\frac{s_0 M_\textup{P}}{\rho_c}\frac{\Gamma_\chi}{m_\chi}\frac{2^{2+\nu}}{\eta}\left(\frac{T_*}{m_\chi}\right)^\nu
   \Gamma\left(\frac{5+\nu}{2}\right)\Gamma\left(\frac{3+\nu}{2}\right) \label{DMeta} \\
  &\simeq & 0.1188 \left(\frac{106,7}{g_*}\right)^{3/2}\frac{\sum_\alpha |y_{\alpha\chi}|^2}{10^{-58}}\, \Pi\,,
  \end{eqnarray}
where $M_\textup{P}$ is the (reduced) Planck mass and
 \begin{equation}\label{Phi}
  \Pi\equiv 10^{-33}\, \frac{2^{\nu}}{7.5 \eta}\left(\frac{T_*}{m_\chi}\right)^\nu
   \frac{\Gamma\left(\frac{5+\nu}{2}\right)\Gamma\left(\frac{3+\nu}{2}\right)}{\Gamma\left(\frac{5}{2}\right)\Gamma\left(\frac{3}{2}\right)}\,,
 \end{equation}
with $\Gamma\left(\frac{5}{2}\right)\Gamma\left(\frac{3}{2}\right)=\frac{3\pi}{2}$. In deriving Eq.\ \eqref{DMeta}, the relation $\displaystyle{\int_0^\infty dx x^{3+\nu}K_1(x)=2^{2+\nu}\Gamma\left(\frac{5+\nu}{2}\right)\Gamma\left(\frac{3+\nu}{2}\right)}$ was used, where $\Gamma(z)$ is the Gamma function. The DM relic abundance and the IceCube data can be coherently explained provided that $\sum_\alpha |y_{\alpha\chi}|^2 \sim 10^{-58}$ and
 \begin{equation}\label{Phivalue}
   \Pi\simeq 1\,,
 \end{equation}
so that one obtains $\Omega_\textup{DM}h^2 \sim 0.1188$, which is consistent with the observed DM abundance \cite{Planck}
 \begin{equation}\label{DM2}
   \Omega_\textup{DM}h^2\Big|_{obs.}=0.1188\pm 0.0010\,.
 \end{equation}

To conclude this section, let us discuss why the Standard Cosmological Model fails to explain the DM relic abundance and the IceCube signal, and one has to invoke a modified cosmology, assuming the minimal model of DM decay described by Eq.\ \eqref{4dop}. In the standard cosmology of GR, one has
 \begin{equation}\label{DMinvdec}
   \Omega_\textup{DM}h^2|_{inv.dec.}=0.1188 \left(\frac{106.75}{g_*}\right)^{3/2}\frac{\sum_\alpha |y_{\alpha \chi}|^2}{7.5\cdot 10^{-25}}\,.
 \end{equation}
From Eq.\ \eqref{DMinvdec} follows that, in order to obtain the correct DM relic abundance \eqref{DM2}, one has to require
 \begin{equation}\label{ychi}
   \sum_{\alpha=e, \mu,\tau}|y_{\alpha\chi}|^2=7.5\cdot 10^{-25}\,.
 \end{equation}
Such a value, in Eq.\ \eqref{DMinvdec}, is in conflict with the value of $\sum_{\alpha=e, \mu,\tau}|y_{\alpha\chi}|^2$ needed to explain the IceCube data. We note, in fact, that: 1) the DM lifetime $\tau_\chi=\Gamma_\chi^{-1}$ has to be larger than the age of the Universe, hence $\tau_\chi> t_U\simeq 4.35\cdot 10^{17}$\,s; 2) the IceCube spectrum provides a constraint on the lower bound for the DM lifetime given by $\tau_\chi^b\simeq 10^{28}$\,s, so that $\tau_\chi \gtrsim \tau_\chi^b$ -- and such a value is, with good approximation,  model-independent \cite{merle}. The combination of Eqs.\ \eqref{ychi} and \eqref{ratechi} then implies
 \[
 \Gamma_\chi\simeq 4.5 \cdot 10^4 \frac{m_\chi}{1\text{PeV}}\,\text{s}^{-1} \qquad \Rightarrow \qquad
 \tau_\chi \simeq 2.2\cdot 10^{-5} \frac{1\text{PeV}}{m_\chi}\,\text{s}\ll t_U\,.
 \]
The IceCube observations require a DM lifetime $\tau_\chi\simeq 10^{28}$\,s that, compared with \eqref{ratechi} for $m_\chi=1$\,PeV, implies
 \begin{equation}\label{yIceCube}
   \sum_{\alpha}| y_{\alpha\chi}|^2\simeq 10^{-58}\,.
 \end{equation}
This value is $33$ orders of magnitude smaller than the value of $\sum_{\alpha=e,\mu, \tau}|y_{\alpha\chi}|^2\sim 10^{-25}$ needed to explain the DM relic abundance (see Eq.\ \eqref{ychi}). Therefore, in the framework provided by the standard cosmology of GR, the IceCube high-energy events and the DM relic abundance are incompatible with the DM production described by Eq.\ \eqref{4dop}. This result holds also for the KM3NeT observation. As we will see below, high-energy neutrino observations can, on one hand, be framed in the context of scalar-tensor gravity cosmology and, on the other hand, be a probe for non-local contributions in the gravitational interaction.

\begin{figure}[bt]
  \centering
  \includegraphics[width=8.5cm]{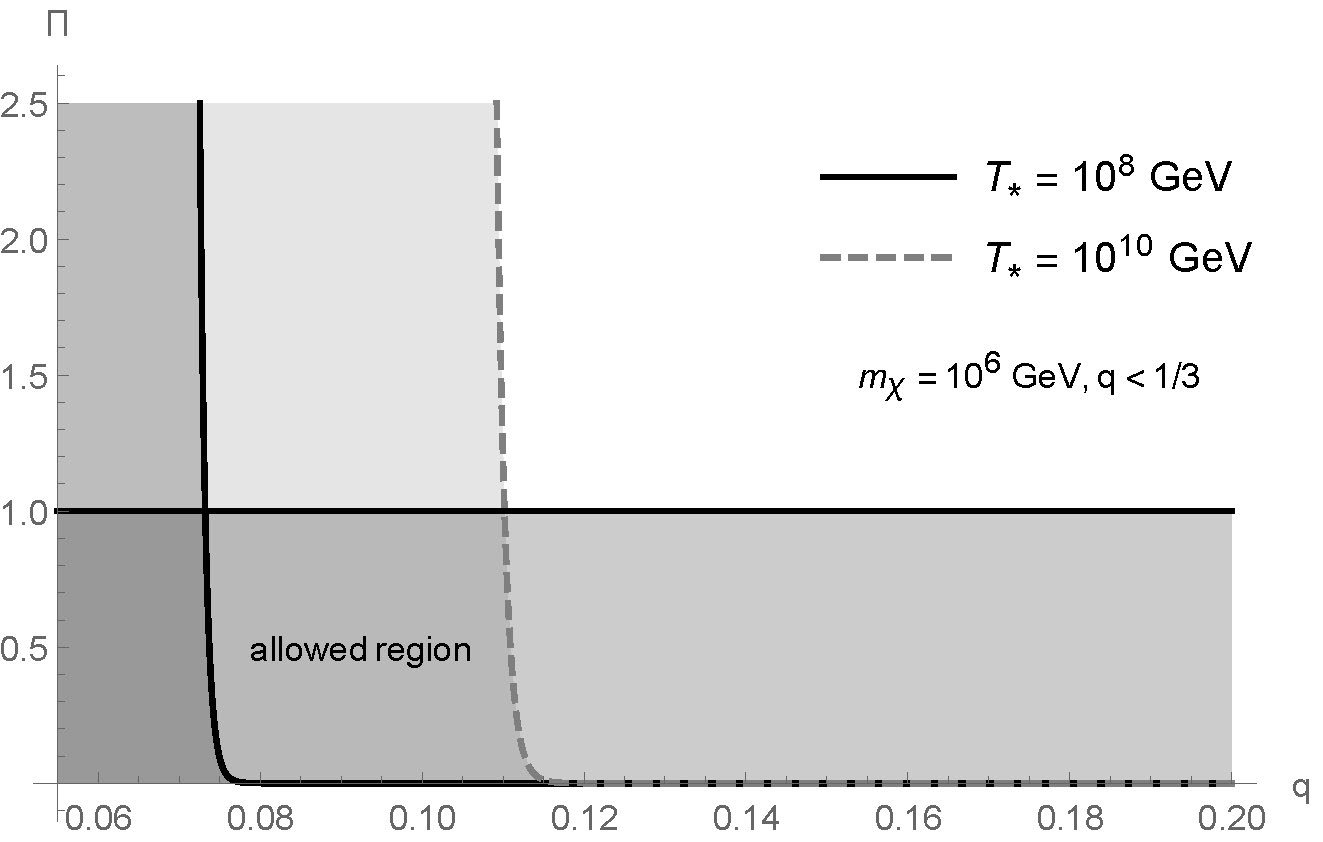}
   \includegraphics[width=8.5cm]{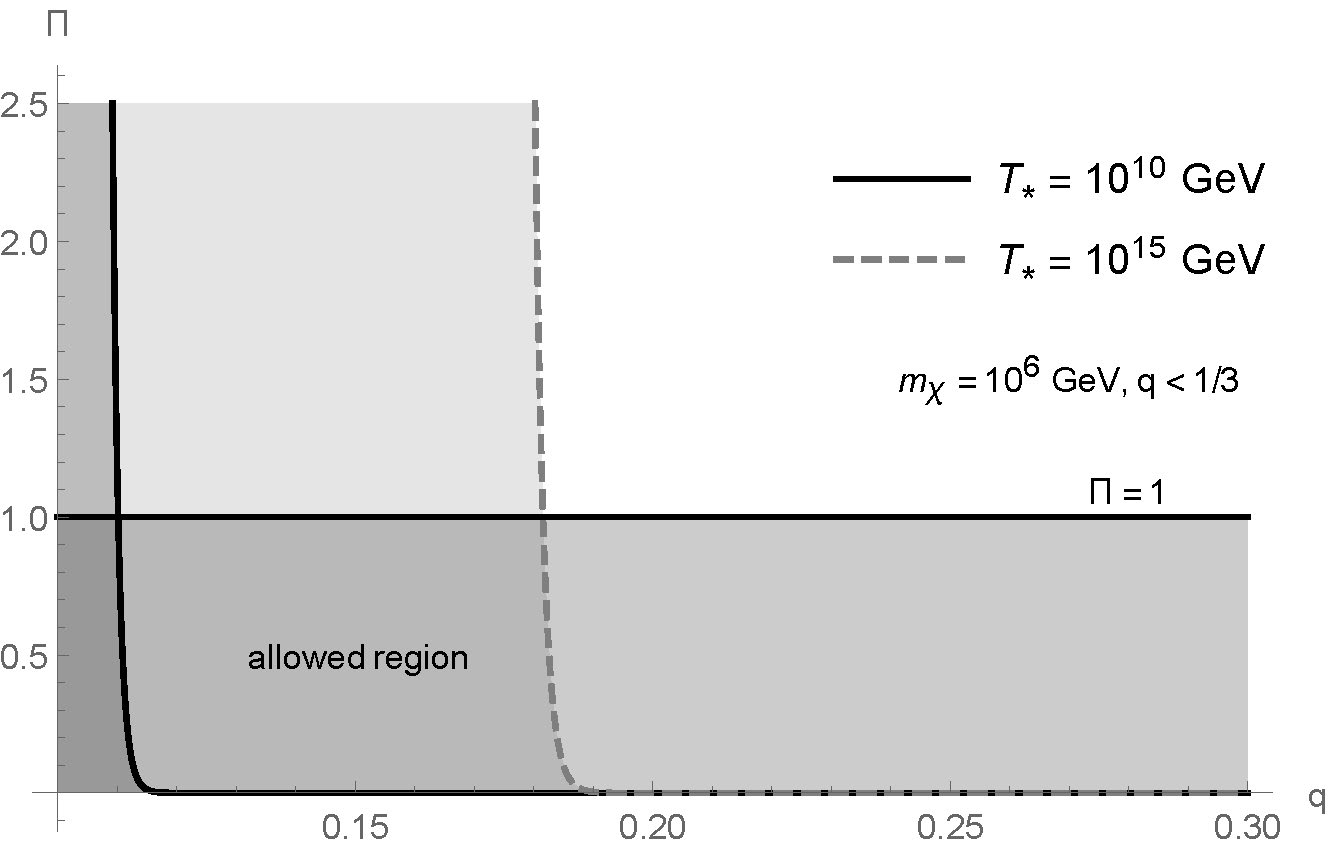}\\
  \caption{The parameter space $\Pi$ vs $q$, with $\Pi$ given by Eq.\ \eqref{Phi} for non-local cosmological models based on curvature and torsion invariants respectively. The allowed region corresponds to $\Pi\lesssim 1$, with the upper bound given by Eq.\ \eqref{Phivalue}. The DM mass is fixed at $m_\chi=1$\,PeV, while the scale factor of the Universe evolves as $a(t)\sim t^q$ with $q<1/3$.}\label{Figp}
\end{figure}

\begin{figure}[bt]
  \centering
  \includegraphics[width=8.5cm]{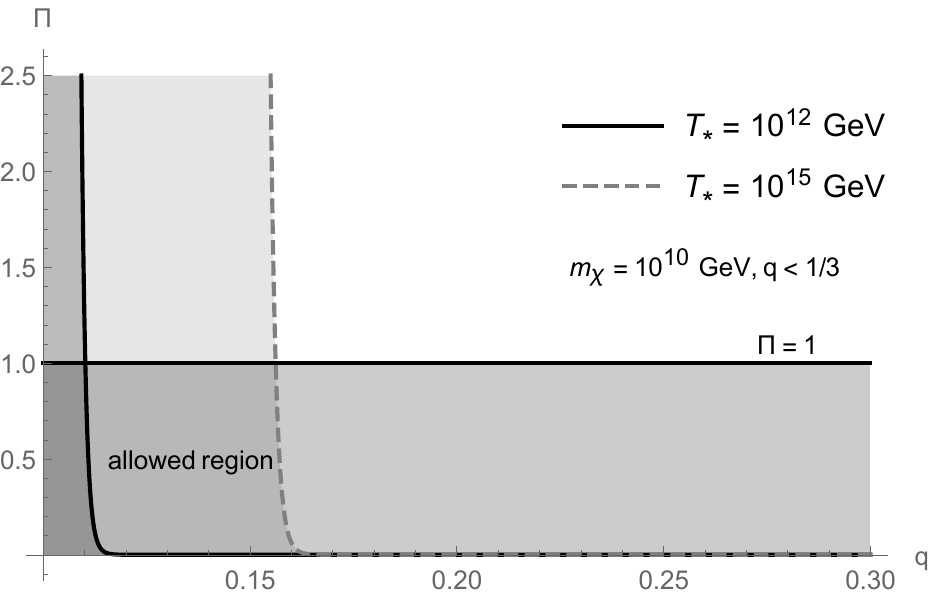}
  \caption{ Same plot as Fig.\ \ref{Figp}, but  with the DM mass set as $m_\chi=10^2$\,PeV.}\label{Figp120}
\end{figure}

 Let us shortly discuss also the KM3NeT/ARCA 120 PeV event. We now set $m_\chi=10^2$\,PeV. According to what was discussed in the previous sections, the IceCube spectrum provides the lower bound $\tau_\chi^b\simeq 10^{28}$\,s of DM lifetime, i.e.\ $\tau_\chi \gtrsim \tau_\chi^b$ \cite{merle}. Inserting Eq.\ \eqref{ychi} into Eq.\ \eqref{ratechi} one obtains $\Gamma_\chi\simeq 4.5 \cdot 10^4 \frac{m_\chi}{\text{1 PeV}}\,\text{s}^{-1} \to \tau_\chi \simeq 2.2\cdot 10^{-7}\,\text{s}\ll t_U$ for $m_\chi\sim 10^2$\,PeV. IceCube observations require the DM decay lifetime to be $\tau_\chi= \sim 10^{28}$\,s, which implies $\sum_{\alpha}| y_{\alpha\chi}|^2\simeq 10^{-60}$. This is about 35 order of magnitudes smaller than the value $\sum_{\alpha=e,\mu, \tau}|y_{\alpha\chi}|^2\sim 10^{-25}$ needed to explain the DM relic abundance (see Eq.\ \eqref{ychi}). Fig.\ \ref{Figp120} shows $\Pi<1$ corresponding to the KM3NeT/ARCA 120 PeV event. As we can see, for such UHE neutrinos the parameters remain compatible with the required constraints $q<1/3$ and $T_* \gtrsim 10^2$\,PeV inferred from the relic abundance.

\section{Discussion and conclusions}
In what follows, we analyse the parameter space of the cosmological models discussed in Sec.\ 2. The viable values of the scale factor $a\sim t^q$ are reported in Table \ref{tab1}. Note that the present parameter $q$ coincides with the parameter $p$ used in Sec.\ 2 for the Noether analysis.

\begin{table}[ht] 
\centering      
\begin{tabular}{l c c }   \hline\hline                
Cosmological model  & p=q values &  \\ [0.1ex] \hline
2.2: Curvature non-local  gravity  & & $q\neq 1/3$      \\ 
2.3: Teleparallel non-local gravity  & & $q\neq 1/3$    \\
2.4: Teleparallel non-local gravity with a boundary term & & $q<1/3$    \\ 
2.5: Horndeski gravity & $\displaystyle{q=\frac{\alpha_1+2\xi_1}{3\xi_1}}$ & $q$ generic   \\ [1ex] 
\hline\hline     
\end{tabular} 
\caption{Values of the exponent $q$ of the scale factor $a\sim t^q$} for various non-local cosmological models.
\label{tab1}
\end{table}
To search for the values of $q$ such that $\Pi\sim 1$ is fulfilled, we parametrise the temperature $T$ as
\begin{equation}\label{Talpha}
  T_* = 10^\alpha \,\text{GeV}\,
\end{equation}
with $\alpha > 6$, which means that we consider temperatures $T> m_\chi$ greater than the DM mass (and therefore greater than the BBN temperature $T_\textup{BBN}\sim 1$\,MeV, as pointed out before). We fix $\alpha$ in the range $8\lesssim  \alpha \lesssim 15$, which means that the transition temperature occurs at scales larger than the DM mass (as required by the freeze-in) until the GUT scale. The parameter space $\Pi$ vs $q$ is plotted in Fig.\ \ref{Figp} for fixed values of the DM mass $m_\chi \sim 1$\,PeV, together with the transition temperatures $T_* \in [10^8,10^{15}]$\,GeV. The allowed region, corresponding to $\Pi \lesssim 1$ (see Eqs.\ \eqref{Phi} and \eqref{Phivalue}), gives $0.08 \lesssim q \lesssim 0.18$. These results show that, before the BBN ends, a significant deviation from the standard evolution of the Universe ($q=0.5$) is needed to account for high-energy neutrinos at PeV scales and the DM relic abundance.

We also note that solutions with a scale factor $a\sim t^q$ with $q>1/3$ are not allowed since they require a transition temperature lower than the PeV energy, a condition excluded by the model under consideration (as the scattering processes would not be kinematically allowed). This result is compatible with those cosmological models that admit scale factor solutions, inferred via the Noether Symmetry Approach, satisfying the constraint $q<1/3$, as reported in Table \ref{tab1}. Moreover, the results we derived are consistent with the constraint $q>-1$, as follows from Eq.\ \eqref{Phi} given that $\Gamma\left(\frac{\nu+3}{2}\right)$ is defined for $\nu>-3$.

 Furthermore, the effective value of $q$ found in our analysis can be recast, in a standard barotropic fluid analogy (where $a(t)\propto t^{2/[3(w+1)]}$), in terms of a perfect fluid with a highly stiff equation of state ($w\gg1$). That said, one of the main features of our study is the fact that all matter, be it ordinary or not, is sub-dominant in the non-local phase preceding the BBN. But since the condition $w\gg1$ characterises the dominant effective geometric fluid rather than the propagation speed of physical matter perturbations, it does not imply superluminal acoustic propagation for physical substances. Instead, it purely encodes a precise form for the amplification factor modifying the expansion rate of the Universe (Eq.\ \eqref{H=AHGRIce}). Therefore, none of the proposed non-local cosmological eras is subject to any causality violations. This fact is supported by the Horndeski-like reformulation of each model: Horndeski gravity represents the most complete scalar-tensor theory containing second-order derivatives in the action while strictly preserving second-order equations of motion too, which protects the scalarised description from any Ostrogradsky instabilities and ghosts.

Another crucial aspect of the proposed non-local models of cosmology to be discussed is their dynamical stability. In the Appendix B we provide a detailed study for the teleparallel model of Eqs.\ \eqref{eq:teleparallel1}--\eqref{eq:teleparallel4}. We chose to focus on this model because it is the only one, out of all those presented in Sec.\ 2, whose stability analysis can easily be carried out analytically -- while the others may require a numerical approach given the much higher complexity of their minisuperspaces. The result of the stability analysis for this non-local model points at a the existence of an unstable repeller as a solution for any positive value of the exponent of the power-law scale factor; this means that, while the early Universe might have given rise to the right conditions to enter the teleparallel phase, and thus produce the observed signatures of UHE neutrinos, this era cannot have lasted indefinitely and indeed must have ended with a ``graceful exit'' at some point. This conclusion corroborates the assumption of the BBN era being not only distinct from the non-local one but also inevitable to follow up. We expect the very same qualitative conclusion to hold true also for the other non-local cosmological models of Sec.\ 2, at least in some special cases thereof (for instance, given the choice of the functions $G_i$ for the Horndeski theory in Eqs.\ \eqref{eq1}--\eqref{eq22}).

In conclusion, to reconcile the current bound on the DM relic abundance with the IceCube data in terms of the four-dimensional operator ${\cal L}_{d=4}=y_{\alpha\chi}\bar L_\alpha H\chi$, we considered four distinct models of non-local cosmology. In particular, we studied the cosmological solutions obtained by applying the Noether Symmetry Approach, focusing on power law solutions. We showed that these dynamical systems representing the non-local evolution of the Universe are able to account for the IceCube observations of high-energy neutrinos as well as for the DM relic abundance observed today in a minimal particle physics model. The main idea we employed relies on the fact that, in non-local cosmology, the expansion rate of the Universe can be expressed in the form $H(T)=A(T)H_{GR}(T)$, encoding in $A(T)$ all the parameters characterising a specific cosmological model. As a consequence, also the thermal history of particles is modified. We then solved the Boltzmann equation to compute the abundance of DM particles and analysed the results for each non-local cosmological model to show that they are consistent with observations for some specific values of the parameter space -- whilst they are \emph{not} consistent in the case of GR. This conclusion could represent an important signature for the non-local nature of the gravitational interaction at cosmological scales.

Finally, it is worth mentioning that a high-energy neutrino event has recently been observed by KM3NeT/ARCA collaboration, the deep-sea neutrino telescope in the Mediterranean Sea. The latter detected a track-like event compatible with an UHE muon with an estimated energy of $120$\,PeV, produced by a neutrino with an even higher energy \cite{Km3}. This is the most energetic neutrino event ever detected, larger than that of any neutrino detected so far. The origin of such extremely energetic neutrino is unclear and, probably, this neutrino might have originated in a cosmic accelerator that is different from those of the common lower-energy neutrinos, or it might represent the first detection of a cosmogenic neutrino (meaning it is the result of the interactions of UHE cosmic rays with the background photons of the Universe). In this case, the energy is $2$ orders of magnitude larger than the typical energy scales observed by the IceCube collaboration. Therefore, the model described here -- that is a minimal DM coupling to Standard Model particles via the four-dimensional operator \eqref{4dop} in the framework of non-local cosmology --, could suggest a novel scenario to account for the KM3NeT/ARCA observation too. These insights open the window to further investigations of non-local cosmologies in the context of astroparticle physics.

\appendix

\section{DM relic abundance in non-local cosmology}
In this appendix we discuss in detail the inverse decay and Yukawa production processes, as well as the fact that the latter is sub-dominant with respect to the former. The dominant contributions to DM production related to the four-dimensional operator \eqref{4dop} are
\begin{itemize}
    \item[-] The {\it inverse decay} processes $\nu_\alpha+H^0\to \chi$ and $l_\alpha + H^+\to \chi$, which occur when $m_\chi> m_H+m_{\nu, l}$. The scattering process is proportional to the factor $|y_{\alpha \chi}|^2$. 
    \item[-] The {\it Yukawa production} processes, such as $t+{\bar t}\to {\bar \nu}_\alpha+\chi$. In this case, the scattering process is proportional to  $|y_{\alpha \chi} y_t|^2$, where $t$ stands for the top quark.
\end{itemize}
The DM relic abundance described by the four-dimensional operator \eqref{4dop} is given by
 \begin{equation}
 \frac{dY_\chi}{dT}=\frac{dY_\chi}{dT}\Big|_{inv.dec.}+\frac{dY_\chi}{dT}\Big|_{Yuk.prod.}\,,
 \end{equation}
with explicit expressions being
 \begin{eqnarray}
   \frac{dY_\chi}{dT}\Big|_{inv.dec.} &=& -\frac{m_\chi^2 \Gamma_\chi}{\pi^2 Hs}\, K_1\left(\frac{m_\chi}{T}\right)\,,\\
   \frac{dY_\chi}{dT}\Big|_{Yuk.prod.} &=& -\frac{1}{512\pi^6Hs}I_{Yuk.prod.}\left(\frac{m_\chi}{T}\right) \,,\\
   I_{Yuk.prod.}\left(\frac{m_\chi}{T}\right) &\equiv &\int d{\tilde s}d\Omega \sum_\alpha
   \frac{W_{t{\bar t}\to{\bar \nu}_\alpha \chi}+2W_{t\nu_{\alpha}\to t\chi}}{\sqrt{\tilde s}}K_1\left(\frac{\sqrt{\tilde s}}{T}\right)\,, \label{dYYukprod}
 \end{eqnarray}
where ${\tilde s}$ is the centre-of-mass energy, $\Gamma_\chi$ is the interaction rate given by
 \begin{equation}
   \Gamma_\chi=\sum_\alpha \frac{|y_{\alpha\chi}|^2}{8\pi}\, m_\chi\,, \qquad \alpha=e, \mu, \tau
 \end{equation}
and $K_1(x)$ is the modified Bessel function of the second kind. According to Eq.\ \eqref{DM1}, one finds
 \begin{eqnarray}\label{DM1G}
   \Omega_\textup{DM}h^2 &=& \frac{2 m_\chi^2 s_0 h^2}{\rho_c}
   \int_0^\infty\frac{dx}{x^2}\left(-\frac{dY_\chi}{dT}\Big|_{T=\frac{m_\chi}{x}}\right)\\
   &=& \frac{2 m_\chi^2 s_0 h^2}{\rho_c}
   \int_0^\infty\frac{dx}{x^2}\frac{1}{A(T)}\left(-\frac{dY_\chi}{dT}\Big|_{T=\frac{m_\chi}{x}}\right)_{GR}\\
   &=& \frac{2 m_\chi^2 s_0 h^2}{\rho_c}
   \int_0^\infty\frac{dx}{x^2}\frac{1}{\pi^2 A(T) H_{GR} s} \left[
  m_\chi^2 \Gamma_\chi K_1 +\frac{1}{512 \pi^6}\, I_{Yuk.prod.}\right]\,, 
 \end{eqnarray}
where we used $H(T)=A(T) H_{GR}(T)$. In the case of standard cosmology, $A(T)=1$, it was show in \cite{merle} (see Fig.\ 4) that for the range of values of $y_t\in [0.5,1]$ (such values of $y_t$ cover all possible values obtained by letting its value run with the energy) and $m_\chi\in [10^4,10^8]$\,GeV, one gets $\Omega_{DM}h^2|_{Yuk.prod.}\simeq 10^{-2}\Omega_{DM}h^2|_{inv.dec.}$, hence $\Omega_{DM}h^2|_{obs.}\simeq \Omega_{DM}h^2|_{inv.dec.}$, i.e.\ the DM relic abundance is mainly generated by inverse decay processes. In a non-local cosmology, $A(T)\neq 1$ (Eq.\ \eqref{A(T)Ice}), the amplification factor $A(T)$ factorises, so that it does not affect the standard result for which $\frac{dY_\chi}{dT}\Big|_{inv.dec.}$ is dominant with respect to $\frac{dY_\chi}{dT}\Big|_{Yuk.prod.}$ (notice that in the parameter space we are considering, $T>T_*$ and $q<1/3$, it follows from Eq.\ \eqref{A(T)Ice} that $A(T)>1$). Since both $(\Omega h^2)_{Yuk.prod.}$ and $(\Omega h^2)_{inv.dec.}$ scale proportionally due to $A(T)$, the ratio $(\Omega h^2)_{Yuk.prod.}/(\Omega h^2)_{inv.dec.}$ is still less than 1, like for standard cosmology.

\section{Stability analysis for the teleparallel non-local cosmological model}
This appendix presents the stability analysis for the cosmological model in Eqs.\ \eqref{eq:teleparallel1}--\eqref{eq:teleparallel4}, which was derived from teleparallel non-local gravity and the Noether Symmetry Approach. The Euler--Lagrange equations for $\theta$, those for $\phi$ and the Hamiltonian constraint ($\mathcal{H}=0$) are respectively
\begin{eqnarray}
\ddot\phi+3H\dot\phi+6H^2&=&0\,,\label{eom1}\\
\ddot\theta+3H\dot\theta-6H^2\frac{c_3}{c_1}f(\phi)&=&0\,,\label{eom2}\\
6H^2[f(\phi)-\theta-1]+\dot\theta\dot\phi&=&0\,.\label{eom3}
\end{eqnarray}
The time derivative of Eq.\ \eqref{eom3} ($\dot{\mathcal{H}}=0$) yields
\begin{equation}\label{eq:time-H}
\biggl(\frac{\dot H}{H}+3H\biggr)\dot\theta\dot\phi-6\frac{c_3}{c_1}H^2f(\phi)\dot\phi+6H^2\dot\theta=0\,.
\end{equation}
To reduce the dynamical system to a simpler form, we use the number of e-folds $N\equiv\log a$, whereby $d/dt=Hd/dN$ (the derivative with respect to $N$ will be denoted by a prime $'$); also, we redefine the dynamical variables as follows:
\begin{equation}
x\equiv\frac{\dot\phi}{H},\qquad y\equiv\frac{\dot\theta}{H},\qquad z\equiv\frac{\frac{c_3}{c_1}f(\phi)}{y}\,.
\end{equation}
Eq.\ \eqref{eq:time-H} then becomes
\begin{equation}
\frac{\dot H}{H^2}=\frac{H'}{H}=6\biggl(z-\frac{1}{x}\biggr)-3\,,
\end{equation}
which can be plugged back into the equations of motion \eqref{eom1} and \eqref{eom2} to obtain a first-order system of coupled differential equations:
\begin{eqnarray}
x'&=&-6xz\,,\\
z'&=&\biggl(\frac{c_3}{c_1}x-\frac{6}{x}\biggr)z\,.
\end{eqnarray}
By doing so, we have effectively managed to reduce the dynamics of the cosmological model to only two degrees of freedom. The equilibrium condition $x'=z'=0$ provides the critical point $(x_c,z_c)=\bigl(\frac{6p}{1-3p},0\bigr)$, hence the Jacobian of the dynamical system gives
\begin{equation}
J(x,z)=
\begin{pmatrix}
-6z&-6x\\\frac{c_3}{c_1}z+\frac{6z}{x^2}&\frac{c_3}{c_1}x-\frac{6}{x}
\end{pmatrix},\qquad
J(x_c,z_c)=
\begin{pmatrix}
0&\frac{36p}{3p-1}\\0&\frac{1}{p}
\end{pmatrix}.
\end{equation}
The eigenvalues of $J(x_c,z_c)$, 0 and $1/p$, determine a condition of unstable repeller for the dynamical system at $(x_c,z_c)$ because the allowed parametric region discussed in Sec.\ 4 requires $p>0$. This conclusion implies that, whilst the Universe could have gone through a non-local teleparallel phase to produce the observed neutrino signatures, this is not an eternal attractor state: the Universe would then naturally evolve away from this cosmic era, which validates the hypothesis of a transient non-local teleparallel phase prior to the BBN.

\acknowledgments
The Authors acknowledge the support of Istituto Nazionale di Fisica Nucleare (INFN), Sez.\ di Napoli, Iniziative Specifiche QGSKY and MoonLight-2. SC is grateful to the {\it Gruppo Nazionale di Fisica Matematica} (GNFM) of {\it Istituto Nazionale di Alta Matematica} (INDAM) for support. This publication is based upon work from COST Action CA21136 -- ``Addressing observational tensions in cosmology with systematics and fundamental physics (CosmoVerse)'', supported by COST (European Cooperation in Science and Technology).

\end{document}